\documentclass[twocolumn,showpacs,preprintnumbers,amsmath,amssymb,floatfix]{revtex4}
\usepackage{epsf}
 \newcommand{\insertplot}[5]{\begin{figure}
 \hfill\hbox to 0.05in{\vbox to #5in{\vfill
 \inputplot{#1}{#4}{#5}}\hfill}
 \hfill\vspace{-.1in}
 \caption{#2}\label{#3}
 \end{figure}}
 \newcommand{\inputplot}[3]{
 \special{ps: plotfile #1}
}

\usepackage[german, english]{babel}
\usepackage{ifthen}
\usepackage{epsfig}
\newcounter{fig}   

\newcommand{\vphi}{\varphi}
\newcommand{\Tr}{{\rm Tr}}

\begin{document}
\title{Monopoles, Antimonopoles and Vortex Rings}
\author{
{\bf Burkhard Kleihaus}
}
\affiliation{
{Department of Mathematical Physics, University College, Dublin,
Belfield, Dublin 4, Ireland}
}
\author{
{\bf Jutta Kunz and Yasha Shnir}
}
\affiliation{
{Fachbereich Physik, Universit\"at Oldenburg, 
D-26111 Oldenburg, Germany}
}
\date{\today}

\begin{abstract}
We present a new class of static axially symmetric solutions 
of SU(2) Yang-Mills-Higgs theory,
where the Higgs field vanishes on rings 
centered around the symmetry axis. 
Associating a magnetic dipole moment with each Higgs vortex ring,
the dipole moments add
for solutions in the trivial topological sector,
whereas they cancel for magnetically charged solutions.
\end{abstract}

\maketitle

{\sl Introduction}
Defects, classical solutions of spontaneously broken gauge theories,
where the Higgs field vanishes at points, lines or surfaces,
are relevant in particle physics and cosmology.
Monopoles, for instance, represent zero-dimensional defects,
vortex solutions or strings are associated with one-dimensional defects,
domain walls represent two-dimensional defects.

Here we present new classical solutions
of SU(2) Yang-Mills-Higgs (YMH) theory with the Higgs field
in the adjoint representation,
where the Higgs field vanishes either at discrete points,
as in single monopoles,
or at rings, as in vortex loops,
or at rings and at a point.

Configuration space of YMH theory consists of sectors,
characterized by the topological charge of the Higgs field.
The 't~Hooft-Polyakov monopole \cite{mono} carries
unit topological charge and possesses spherical symmetry. 
Multimonopoles with higher topological charge 
possess at most axial symmetry 
\cite{WeinbergGuth,RebbiRossi,mmono},
or no rotational symmetry at all \cite{monoDS}.
The magnetic charge of the (multi-)monopoles is proportional
to their topological charge.

In the Bogomol'nyi-Prasad-Sommerfield (BPS) limit 
of vanishing Higgs potential
monopoles and multi-monopoles are obtained as 
solutions of the first order Bogomol'nyi equations \cite{Bogo}.
The energy of these solutions satisfies exactly the lower energy 
bound given by the topological charge. 
Since the repulsive and attractive forces between monopoles
exactly compensate, monopoles experience no net interaction.

As shown by Taubes \cite{Taubes}, each topological sector 
contains further smooth, finite energy solutions,
which do not satisfy the Bogomol'nyi equations, 
but only the second order Euler-Lagrange equations.
These solutions form saddlepoints of the energy functional. 
Their energy exceeds the Bogomol'nyi bound. 

In the topologically trivial sector the simplest such solution 
is axially symmetric, and corresponds to an equilibrium state of a 
monopole-antimonopole pair \cite{Rueber,mapKK}.
Here the forces acting on the monopole and antimonopole
are balanced, resulting in this (unstable) state, 
which carries an Abelian magnetic dipole moment. The Abelian magnetic field
resembles the field of a physical dipole with magnetic charges
localized on the symmetry axis at the equilibrium distance.

Recently, more general static equilibrium solutions 
have been constructed,
representing chains, where $m$ monopoles and antimonopoles alternate
along the symmetry axis \cite{KKS}.
$m$-chains in the topologically trivial sector
carry a magnetic dipole moment and no charge,
whereas $m$-chains in the sector with topological charge one
carry charge and no magnetic dipole moment \cite{KKS}.

Here we address the question, whether chains of 
$m$ multimonopoles and antimonopoles,
each with charge $n$, can exist in static equilibrium.
The simplest such generalization, an equilibrium state of a charge $2$ 
monopole and a charge $-2$ antimonopole 
was obtained recently \cite{Tigran}. 
We find that, beyond charge $n=2$, no such equilibrium configurations
of localized point charges are possible. 
Instead a new type of equilibrium solution appears.

{\sl Ansatz and boundary conditions}
We consider SU(2) YMH theory in the BPS limit,
\begin{equation}
-{\cal L} = \frac{1}{2} \Tr\left( F_{\mu\nu} F^{\mu\nu}\right)
                +\frac{1}{4} \Tr\left( D_\mu \Phi D^\mu \Phi \right)
	 \ ,
\nonumber
\end{equation}
with su(2) gauge potential $A_\mu = A_\mu^a \tau^a/2$,
field strength tensor
$F_{\mu\nu} = \partial_\mu A_\nu - \partial_\nu A_\mu + i  [A_\mu, A_\nu]$,
and covariant derivative of the Higgs field
$D_\mu \Phi = \partial_\mu \Phi +i  [A_\mu, \Phi]$.

Generalizing both the Ansatz for the monopole-antimonopole pairs and chains 
\cite{Rueber,mapKK,Tigran,KKS},
and the axially symmetric multimonopole ansatz \cite{RebbiRossi,KKT},
we parametrize the gauge potential and the Higgs field by 
\begin{eqnarray}
A_\mu dx^\mu
& = &
\left( \frac{K_1}{r} dr + (1-K_2)d\theta\right)\frac{\tau_\vphi^{(n)}}{2}
\nonumber \\
&-&n \sin\theta \left( K_3\frac{\tau_r^{(n,m)}}{2}
                     +(1-K_4)\frac{\tau_\theta^{(n,m)}}{2}\right) d\vphi
\nonumber \\
\Phi
& = &
\Phi_1\tau_r^{(n,m)}+ \Phi_2\tau_\theta^{(n,m)} \  
\nonumber
\end{eqnarray}
with su(2) matrices
$\tau_r^{(n,m)}=
\sin(m\theta) \tau_\rho^{(n)} + \cos(m\theta) \tau_z$,
$\tau_\theta^{(n,m)}=
\cos(m\theta) \tau_\rho^{(n)} - \sin(m\theta) \tau_z$,
$\tau_\vphi^{(n)}=
 -\sin(n\vphi) \tau_x + \cos(n\vphi)\tau_y$
and $\tau_\rho^{(n)} =\cos(n\vphi) \tau_x + \sin(n\vphi)\tau_y $.
We refer to the integers $m$ and $n$ as 
$\theta$ winding number and $\vphi$ winding number, respectively.
The profile functions $K_1$ -- $K_4$ and $\Phi_1$, $\Phi_2$ 
depend on the coordinates $r$ and $\theta$, only.
The ansatz possesses a residual U(1) gauge symmetry.
To fix the gauge we impose
the condition $r\partial_r K_1 - \partial_\theta K_2 = 0$ \cite{KKT,KKS}.

To obtain regular solutions with finite energy and energy density 
we have to impose appropriate boundary conditions. 
Regularity at the origin requires
$K_1=K_3=0$, $K_2=K_4=1$,
$\sin(m\theta) \Phi_1+ \cos(m\theta) \Phi_2 = 0$,
$\partial_r\left[\cos(m\theta) \Phi_1 - \sin(m\theta) \Phi_2\right] = 0$.
At infinity we require 
the solutions in the vacuum sector ($m=2k$) to tend to
a gauge transformed trivial solution, 
$$
\Phi \ \longrightarrow U \tau_z U^\dagger \   , \ \ \
A_\mu \ \longrightarrow  \ i \partial_\mu U U^\dagger \ ,
$$
and the solutions in the topological charge $n$ sector ($m=2k+1$)
to tend to
$$
\Phi  \longrightarrow  U \Phi_\infty^{(1,n)} U^\dagger \   , \ \ \
A_\mu \ \longrightarrow \ U A_{\mu \infty}^{(1,n)} U^\dagger
+i \partial_\mu U U^\dagger \  ,
$$
where
$$ \Phi_\infty^{(1,n)} =\tau_r^{(1,n)}\ , \ \ \
A_{\mu \infty}^{(1,n)}dx^\mu =
\frac{\tau_\vphi^{(n)}}{2} d\theta
- n\sin\theta \frac{\tau_\theta^{(1,n)}}{2} d\vphi
$$
is the asymptotic solution of a charge $n$ multimonopole,
and  $U = \exp\{-i k \theta\tau_\vphi^{(n)}\}$, both
for even and odd $m$.
Consequently, solutions with even $m$ have vanishing magnetic charge,
whereas solutions with odd $m$ possess magnetic charge $n$.

In terms of the functions $K_1 - K_4$, $\Phi_1$, $\Phi_2$ these boundary
conditions read
$K_1 =0$,
$K_2 = 1 - m$,
$K_3 =\left({\cos\theta - \cos(m\theta)}\right)/{\sin\theta}$
for odd $m$ and
$K_3 =\left({1 - \cos(m\theta)}\right)/{\sin\theta}$
for even $m$,
$K_4 =1- \sin(m\theta)/\sin\theta$,
$\Phi_1=1$, and $\Phi_2 =0$.

Regularity on the $z$-axis, finally, requires
$K_1 = K_3 = \Phi_2 =0$,
$\partial_\theta K_2 = \partial_\theta K_4 = \partial_\theta \Phi_1 =0$,
for $\theta = 0$ and $\theta = \pi$.

Defining the Abelian magnetic field via the 't Hooft tensor 
with normalized Higgs field $\hat \Phi$ \cite{mono},
$$
{\cal F}_{\mu\nu} = {\rm Tr} \left\{ \hat \Phi F_{\mu\nu}
- \frac{i}{2} \hat \Phi D_\mu \hat \Phi D_\nu \hat \Phi \right\}
$$
we note, that only solutions with even $m$ 
possess an Abelian magnetic dipole moment \cite{KKS}.

With this Ansatz the general field equations 
reduce to six PDEs in the coordinates $r$ and $\theta$,
which are solved numerically, subject to the above boundary conditions.

{\sl Results}
The $m$-chains constructed in \cite{KKS}
are characterized by $\theta$ winding number $m>1$ 
and $\vphi$ winding number $n=1$.
In these solutions $m$ monopoles and antimonopoles 
are located on the symmetry axis,
with (roughly) equal distance between them.
Their energy increases (approximately)
linearly with $m$, and likewise does the Abelian magnetic moment 
of $m$-chains with even $m$ \cite{KKS}.

Let us now consider chains consisting of multimonopoles
with $\vphi$ winding number $n=2$. 
For such chains with $m\le 5$ the energies, magnetic moments and locations 
of the Higgs zeros are shown in Table 1.
Their energy increases (approximately)
still linearly with $m$, and so does the magnetic moment 
of the chains with even $m$.

Identifying the locations of the Higgs zeros on the symmetry axis
with the locations of the monopoles and antimonopoles, we observe that
when each pole carries charge $n=2$,
the zeros form pairs, when possible,
where the distance between the monopole and the antimonopole of a pair
is less than the distance to the neighboring monopole or antimonopole,
belonging to the next pair.

\begin{widetext}
\parbox{\textwidth}{
\centerline{
\begin{tabular}{|c|cccc|cccc|cccc|}
 \hline
   \multicolumn{1}{|c|}{}
 & \multicolumn{4}{|c|}{$E[4\pi\eta$]}
 &  \multicolumn{4}{|c|}{$\mu/n$} 
 &  \multicolumn{4}{|c|}{($\rho_i,\pm z_i$)} \\
 \hline
$m/n$ &  1   &   2  &   3  &   4  &   1 &   2 &   3 &   4 &
1 &   2 &   3 &   4
 \\
 \hline
1     & 1.00 & 2.00 & 3.00 & 4.00 & 0.0  & 0.0  & 0.0  & 0.0 &
(0, 0) & (0, 0)& (0, 0) & (0, 0)\\
 \hline
2     & 1.70 & 2.96 & 4.03 & 5.01 & 2.36 & 2.38 & 2.6  & 2.87 &
(0, 2.1) & (0, 0.9) & (3.0, 0)& (4.9, 0) \\
 \hline
3     & 2.44 & 4.17 & 5.62 & 6.96 & 0.0  & 0.0  & 0.0  & 0.0 &
\begin{tabular}{c} (0, 0) \\ (0, 4.7) \end{tabular} &
\begin{tabular}{c} (0, 0) \\ (0, 3.2) \end{tabular} &  
\begin{tabular}{c} (0, 0) \\ (2.0, 1.2) \end{tabular} & 
\begin{tabular}{c} (0, 0) \\ (3.6, 0)\\ (4.3, 0.8) \end{tabular}   \\
 \hline
4     & 3.12 & 5.07 & 6.63 & 8.00 & 4.93 & 4.81 & 5.20 & 5.42 &
\begin{tabular}{c} (0, 2.4) \\ (0, 7.0) \end{tabular} & 
\begin{tabular}{c} (0, 2.0) \\ (0, 4.9) \end{tabular} & 
\begin{tabular}{c} (3.0, 3.0) \end{tabular} &
\begin{tabular}{c} (5.4, 2.8) \end{tabular}\\
 \hline
5     & 3.78 & 6.11 & 7.96 & 9.59 & 0.0  & 0.0  & 0.0  & 0.0 &
\begin{tabular}{c} (0, 0) \\ (0, 4.8) \\ (0, 9.6)  \end{tabular}&
\begin{tabular}{c} (0, 0) \\ (0, 4.1) \\ (0, 7.3)  \end{tabular}&
\begin{tabular}{c} (0, 0) \\ (3.1, 5.2)  \end{tabular}&
\begin{tabular}{c} (0, 0) \\ (5.7, 4.7)  \end{tabular}
\\
 \hline
\end{tabular}\vspace{7.mm}}
{\bf Table 1}
The dimensionless energy, the dipole moment per winding 
number $\mu/n$ and the coordinates of the zeros of the
Higgs field are given for several values of $m$ and $n$.\vspace{7.mm}\\
}
\end{widetext}

We observe furthermore,
that the equilibrium distance of the monopole-antimonopole pair
composed of $n=2$ multimonopoles
is smaller than the equilibrium distance of the monopole-antimonopole pair 
composed of single monopoles.
Thus the higher attraction between the poles of a pair with 
charge $n=2$ is balanced by the repulsion only 
at a smaller equilibrium distance.

When increasing the charge of the poles to $n>2$,
we expect this trend to continue. 
The monopoles and antimonopoles of the pairs
should approach each other further,
settling at a still smaller equilibrium distance.

Constructing solutions with charge $n=3$, 
however, we do not find chains at all.
Now there is no longer sufficient repulsion
to balance the strong attraction between 
the 3-monopoles and 3-antimonopoles.
Instead of chains, we now observe solutions with vortex rings,
where the Higgs field vanishes on closed rings around the symmetry axis.

To better understand these findings let us consider
unphysical intermediate configurations, where we allow the
$\vphi$ winding number $n$ to continuously vary between the
physical integer values.
Beginning with the simplest such solution, the $m=2$ solution,
we observe, that the zeros of the solution with winding number $n$
continue to approach each other when $n$ is increased beyond 2,
until they merge at the origin. 
Here the pole and antipole do not annihilate, however.
We conclude, that
this is not allowed by the imposed symmetries and
boundary conditions.
Instead the Higgs zero changes its character completely,
when $n$ is further increased.
It turns into a ring with increasing radius for increasing $n$.
The physical 3-monopole-3-antimonopole solution then
has a single ring of zeros of the Higgs field and no point zeros.
A further increase of $n$ only increases the radius of the ring 
further.

\begin{widetext}
\parbox{\textwidth}{
\begin{center}
\mbox{\hspace{-2.0cm}
\epsfxsize=17.cm
\epsffile{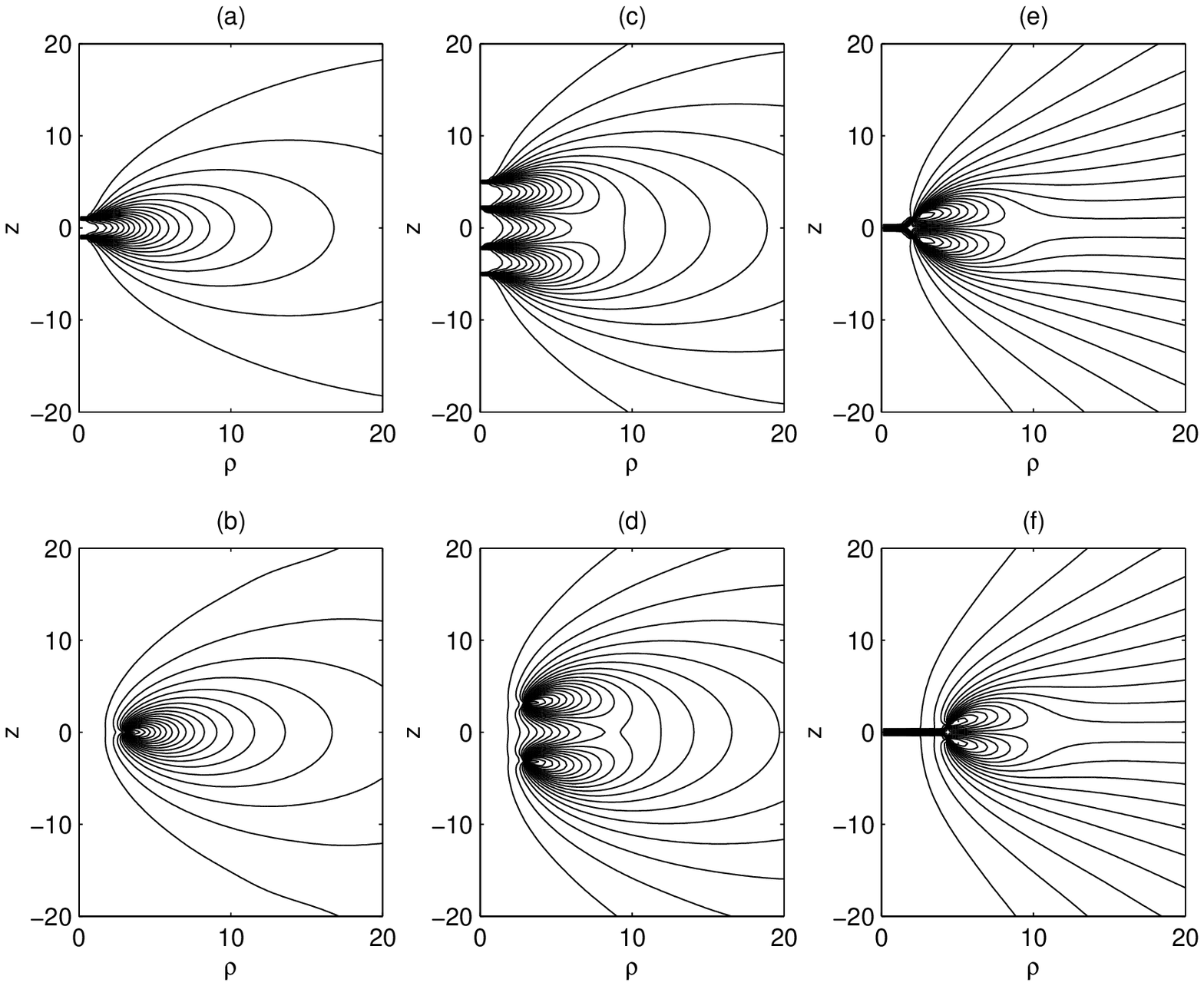}
}
\end{center}
}

{\bf Fig.~1} 
The field lines of the Abelian magnetic field are shown for
$m=2$, $n=3$ (a), $m=2$, $n=3$ (b), 
$m=4$, $n=3$ (c), $m=4$, $n=3$ (d),
$m=3$, $n=3$ (e) and $m=3$, $n=4$ (f).

\vspace{0.5cm}
\end{widetext}

Considering the magnetic moment of the $m=2$ solutions,
we observe, that it is (roughly) proportional to $n$.
The pair of poles on the $z$-axis for $n=2$ clearly 
gives rise to the magnetic dipole moment of a physical dipole,
as illustrated in Fig.~1a, where we show the field lines
of the magnetic field, obtained from the 't Hooft tensor.
As seen in Fig.~1b, the ring of zeros
also gives rise to a magnetic dipole field,
which however looks like the field of a ring of mathematical dipoles.
This corresponds to the simple picture that the positive and negative
charges have merged but not annihilated, and then spread out
on a ring.

The solutions with even $\theta$ winding number
reside in the vacuum sector. 
For $m=2k>2$ solutions it is now clear
how they evolve, when the $\vphi$ winding number is increased beyond $n=2$.
Starting from $k$ pairs of physical dipoles, 
the pairs merge and form $k$ vortex rings, which carry
the dipole strength of the solutions.
This is illustrated in Fig.~1c for $m=4,n=2$, 
and in Fig.~1d for $m=4,n=3$.
As seen in Table 1,
the total dipole moment increases (roughly)
linearly both with $m$ and $n$, 
since there are $m/2$ rings, each formed from charges $\pm n$.

The solutions with odd $\theta$ winding number
reside in the topological sector with charge $n$.
For $m=4k+1$ the situation is somewhat similar to the above.
Here a single $n$-monopole remains at the origin,
whereas all other zeros form pairs, which for $n>2$ approach each other,
merge and form rings carrying dipole strength. 
Since, however, a dipole on the positive axis and its respective
counterpart on the negative axis have opposite orientation, 
their contributions cancel in the total magnetic moment.
Thus the magnetic moment remains zero, as it must, because of symmetry
\cite{KKS}.

For $m=4k-1$, on the other hand, the situation is more complicated,
because in this case there are for $n=2$
always 3 poles on the $z$-axis,
which cannot form pairs, such that all zeros belong to a pair,
symmetrically located around the origin.
For the simplest case, $m=3$, we observe, that
two vortices appear in the charge $n=3$ solution, emerging from
the upper and lower unpaired zero, respectively,
carrying opposite dipole strength. 
For $n=4$ a third ring appears, emerging from the zero at the origin,
which however does not carry a dipole moment.
The magnetic field lines of the $n=3$ and 4 solutions are shown
in Figs.~1e and f, respectively.
For $n=5$ finally all three rings merge to form a single ring.
Further details of these solutions will be given elsewhere
\cite{long}.

Concluding, we have found new static axially symmetric solutions of
SU(2) YMH theory,
characterized by two winding numbers, $m$ and $n$.
For $n \le 2$ the Higgs field vanishes on $m$ discrete points
on the $z$-axis,
for $n>2$ it vanishes on $m/2$ rings centered around
the $z$-axis for even $m$,
while for odd $m$, it vanishes on one or more rings and at the origin.
Solutions with even $m$ reside in the topologically trivial sector.
They carry no magnetic charge but a magnetic dipole moment, 
(roughly) proportional to the product $m$ times $n$.
In contrast, solutions with odd $m$ reside in sectors with 
non-trivial topology. They carry magnetic charge $n$ and possess no
magnetic dipole moment.
Analogous results holds for finite Higgs self-coupling \cite{long}.

We expect that solutions of similar structure might exist in 
Weinberg-Salam theory \cite{BrihayeKunz}, where so far only 
the sphaleron ($m=1,n=1$) \cite{S}, the multisphalerons ($m=1,n=>1$)
\cite{KK}, and the sphaleron $S^*$ ($m=2,n=1$) \cite{SS} are known.

Rings of vanishing or small Higgs field
are also present in Alice electrodynamics, where they
carry magnetic Cheshire charge \cite{Bais},
while closed knotted vortices can arise
in theories, allowing for solutions with
non-trivial Hopf number \cite{Niemi}.


\end{document}